\documentclass[12pt]{article}
\usepackage[margin=.8in]{geometry}
\usepackage{authblk} 
\usepackage{comment} 
\usepackage{fullpage} 
\usepackage{xcolor}
\usepackage{subfiles}

\usepackage{array}
\usepackage{booktabs}
\usepackage{xurl} 
\usepackage{hyperref}
\hypersetup{breaklinks=true, colorlinks=true, urlcolor=blue,citecolor=green} 
\usepackage{algpseudocode}

\usepackage{times}
\usepackage{mathtools}
\usepackage{amsmath}
\usepackage{bm}
\usepackage{amssymb}
\usepackage{pifont}
\usepackage{marginnote}
\usepackage{graphicx}
\usepackage{dcolumn}%
\usepackage{dcolumn}
\usepackage{bm}
\usepackage{mathtools}
\usepackage{color}
\usepackage[linesnumbered,ruled,vlined]{algorithm2e}

\usepackage[font=small,labelfont=bf]{caption}

\makeatletter
\renewcommand{\maketitle}{\bgroup\setlength{\parindent}{0pt}
\begin{flushleft}
 \fontsize{16}{32}
 \textbf{\@title}
 
   \fontsize{12}{24}
  \@author
\end{flushleft}\egroup
}
\makeatother
\newfam\bboardfam
\font\tenbboard=msbm10  
 \font\sevenbboard=msbm7
   \font\fivebboard=msbm5 
\textfont\bboardfam=\tenbboard
\scriptfont\bboardfam=\sevenbboard
\scriptscriptfont\bboardfam=\fivebboard

\title{Active feature selection discovers minimal gene sets for classifying cell types and disease states with single-cell mRNA-seq data}

\author[1]{Xiaoqiao Chen}
\author[2,3]{Sisi Chen}
\author[1,2,3,4]{Matt Thomson}

\affil[1]{Department of Computing and Mathematical Sciences, California Institute of Technology. Pasadena, California, 91125, USA.}

\affil[2]{Division of Biology and Biological Engineering, California Institute of Technology. Pasadena, California, 91125, USA.}

\affil[3]{Beckman Center for Single-cell Profiling and Engineering. Pasadena, California, 91125, USA. }

\affil[4]{correspondence to: mthomson@caltech.edu\vspace{0cm}}

\date{June 2021}

\begin{document}

\maketitle
\section*{Abstract}
Sequencing costs currently prohibit the application of single-cell mRNA-seq to many biological and clinical analyses. Targeted single-cell mRNA-sequencing reduces sequencing costs by profiling reduced gene sets that capture biological information with a minimal number of genes. Here, we introduce an active learning method (ActiveSVM) that identifies minimal but highly-informative gene sets  that enable the identification of  cell-types, physiological states, and genetic perturbations in single-cell data using a small number of genes. Our active feature selection procedure generates minimal gene sets from single-cell data through an iterative cell-type classification task where misclassified cells are examined at each round of analysis to identify maximally informative genes through an `active' support vector machine (ActiveSVM) classifier. By focusing computational resources on misclassified cells, ActiveSVM scales to analyze data sets with over a million single cells. We demonstrate that ActiveSVM feature selection identifies gene sets that enable $~90\%$ cell-type classification accuracy across a variety of data sets including cell atlas and disease characterization data sets. The method generalizes to reveal genes that respond to genetic perturbations and to identify region specific gene expression patterns in spatial transcriptomics data. The discovery of small but highly informative gene sets should enable substantial reductions in the number of measurements necessary for application of single-cell mRNA-seq to clinical tests, therapeutic discovery, and genetic screens. 

\section*{Introduction}
Single-cell mRNA-seq methods have scaled to allow routine transcriptome-scale profiling of thousands of cells per experimental run. While single cell mRNA-seq approaches provide insights into many different biological and biomedical problems, high sequencing costs prohibit the broad application of single-cell mRNA-seq in many exploratory assays such as small molecule and genetic screens, and in cost-sensitive clinical assays. The sequencing bottleneck has led to the development of targeted mRNA-seq strategies that reduce sequencing costs, by up to $90\%$, by focusing sequencing resources on highly informative genes for a given biological question or an analysis \cite{heimberg2016low, fan2015combinatorial,replogle2020combinatorial,marshall2020hypr,sheynkman2020orf,riemondy2019recovery}. Commercial gene-targeting kits, for example, reduce sequencing costs through selective amplification of specific transcripts using $~1000$ gene-targeting primers.

Targeted sequencing approaches require computational methods to identify highly informative genes for specific biological questions, systems, or conditions. A range of computational approaches including differential gene expression analysis and PCA can be applied to identify highly informative genes \cite{heimberg2016low}. However, current methods for defining minimal gene sets are computationally expensive to apply to large single-cell mRNA-seq data sets and often require heuristic user-defined thresholds for gene selection  \cite{Delaney2019-ui, wang2019scmarker}. As an example, computational approaches based upon matrix factorization (PCA\cite{hotelling1933analysis}, NNMF\cite{lee1999learning}), are typically applied to complete data sets and so are computationally intensive when data sets scale into the millions of cells \cite{Song2021.02.09.430550, bhaduri2018identification}. Further, gene set selection after matrix factorization requires heuristic strategies for thresholding coefficients in gene vectors extracted by PCA or NNMF, and then asking whether the selected genes retain core biological information.  

Here,  inspired by active learning\cite{felder2009active} approaches, we develop a computational method that selects minimal gene sets capable of reliably identifying cell-types and transcriptional states in single-cell mRNA-seq.  Our method, ActiveSVM, constructs minimal gene sets by performing an iterative support vector machine classification task \cite{ruckstiess2011sequential,noble2006support}. In ActiveSVM the minimal gene set grows from an initial random seed. At each round, ActiveSVM classifies cells into classes that are provided by unsupervised clustering of cell-states or by used-supplied experimental labels.  The ActiveSVM procedure analyzes cells that are misclassified with the current gene set, and, then, identifies maximally informative genes that are added to the growing gene set to improve classification. Traditional active learning algorithms query an oracle for training examples that meet a criteria \cite{settles2009active}.  Our ActiveSVM procedure actively queries the output of an SVM classifier for cells that classify poorly, and then performs detailed analysis of the specific misclassified cells to select maximally informative genes which are, then, added to a growing gene set. By focusing on a well-defined classification task, we ensure that the gene sets discovered by ActiveSVM retain biological information. 

The central contribution of ActiveSVM is that the method can scale to large single-cell data sets with more than one million cells. We demonstrate, for example, that ActiveSVM can analyze a mouse brain data set with $1.3$ million cells and requires only hours of computational time.  ActiveSVM scales to large data sets because the procedure must only analyze the full-transcriptome of cells that classify poorly with the current gene set.  As the procedure focuses computational resources on poorly classified cells, the method can be applied to large data sets to discover small sets of genes that can distinguish between cell-types at high accuracy. In addition to scaling, the classification paradigm generalizes to a range of single-cell data analysis tasks including the identification of disease markers, genes that respond to Cas9 perturbation , and the identification of region specific genes in spatial transcriptomitcs .

To demonstrate the performance of ActiveSVM, we apply the method to a series of single-cell genomics data sets and analysis tasks. We identify minimal gene sets for cell-state classification in human peripheral blood mononuclear cells (PBMCs) \cite{zheng2017massively}, the megacell mouse brain data set \cite{genomics20171}, and 
the Tabula Muris mouse tissue survey\cite{tabula2018single}. We demonstrate application of ActiveSVM to identify disease markers by analyzing a data set of healthy and multiple myeloma patient PBMCs \cite{chen2020dissecting}. To highlight the generality of the method, we apply ActiveSVM to identify genes impacted by Cas9 based gene-knock down in perturb-seq \cite{dixit2016perturb}. Further, we show that ActiveSVM can identify gene sets that mark specific spatial locations of a tissue through analysis of spatial transcriptomics data \cite{eng2019transcriptome}. To benchmark the method, we compare the performance of the method to six conventional feature selection methods, showing that our method outperforms these methods in classification accuracy. Gene sets constructed by ActiveSVM are both small and highly efficient, for example, classifying human immune cell types within PMBCs using as few as $15$ genes and classifying 55 cell-states in Tabula Muris with $<150$ genes. The gene sets we discover include both classical markers and genes not previously established as canonical cell-state markers. Conceptually, ActiveSVM demonstrates how active sampling strategies can be applied to enable the scaling of algorithms to the large data sets generated single-cell genomics.   

\begin{figure}
    \centering
    \includegraphics[width=\textwidth]{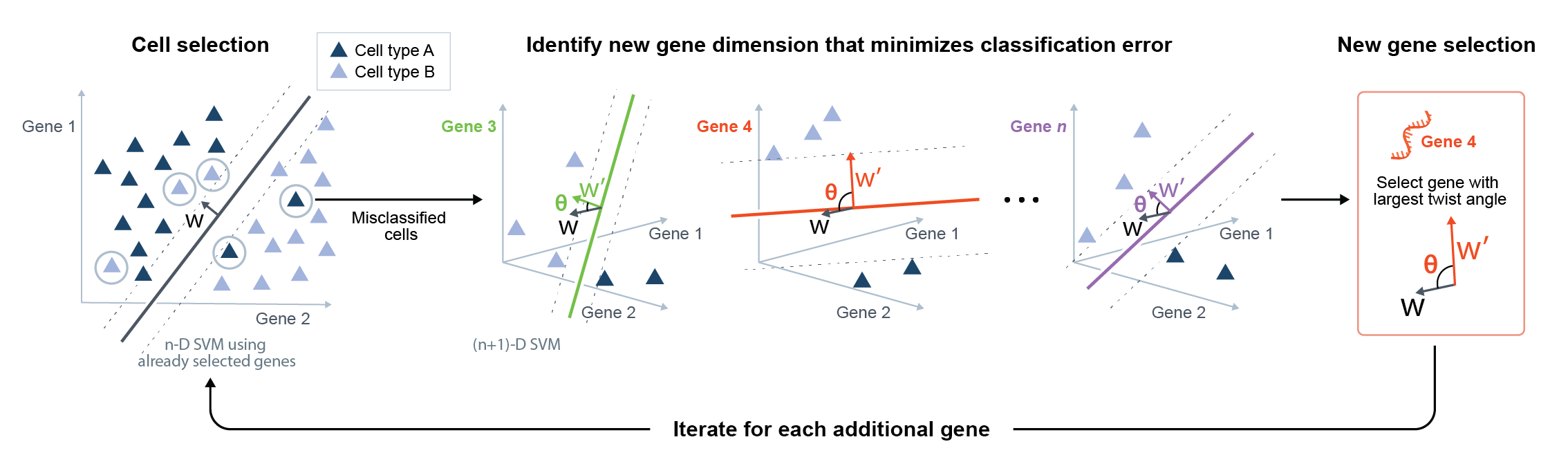}
    \caption{\textbf{Description of ActiveSVM Feature Selection.} At the $n$-th step, an $n$-D SVM using all already selected genes is trained to select a certain number of misclassified cells, which is the cell selection step. In the gene selection step, the least classifiable cells are taken as the training set. Based on this training set, $N$-$n$ ($n$+1)-D SVMs are trained, where $n$ dimensions are the genes already selected and the last dimension is one of the previously unselected candidate genes. Then we would obtain $N$-$n$ weights $w'$ corresponding to $N$-$n$ unselected genes as well as $N$-$n$ margin rotation angle $\theta$ between every $w'$ and the original weight $w$ of the $n$-D SVM. The gene with the maximum rotation of margin is selected for the next round.}
    \label{fig:diagram}
\end{figure}

\section*{Method}

We developed a computational method based on support vector machine (SVM) classifier to identify compact gene sets that distinguish cell-states in single-cell data. In the conventional Sequential Feature Selection (SFS) \cite{ruckstiess2011sequential}, features are selected one-by-one in a greedy strategy to optimize an objective function. Here, we develop an active SVM (ActiveSVM) feature selection method, where we only analyze the subset of incorrectly classified cells at the current step and then select the new gene features based upon those cells. This active learning strategy enables the efficient computation of small gene sets across large data sets by minimizing the total number of cells and genes that are analyzed.

ActiveSVM proceeds through rounds of classification and gene selection based on a set of cell labels. Cell labels can be derived from unsupservised analysis, experimental meta-data, or biological knowledge of cell-type marker genes. A common work-flow in single-cell mRNA-seq experiments defines a series of cell-states or cell-types using unsupervised clustering of cells \cite{wolf2018scanpy,macosko2015highly}. Therefore, we developed our method to accept as input the cell-state labels that are typically derived from unsupervised clustering. We, then,  utilize the cell-state labels to identify a minimal set of marker genes that can retain the separation between cell-states with a minimal set of gene features. We note that our method can also accept user supplied cell-type labels as input if a user seeks to identify new genes that separate cell-states based upon biologically curated markers.  

Our ActiveSVM procedure starts with an empty gene set, an empty cell set and a list of candidate genes and cells. The algorithm iteratively selects genes and classifies cells using identified genes by training a SVM model to classify the cell-types according to labels. The algorithm identifies cells in the data set that classify poorly given the current gene set, and uses misclassified cells to select additional genes to improve classification accuracy on the entire data set. We supply `min-complexity' and `min-cell versions' of ActiveSVM algorithm. The min-complexity algorithm samples a fixed number of misclassified cells and directly uses them as the cell set to select the next gene. The min-cell algorithm re-uses the misclassified cells selected in previous iterations to reduce the total number of required cells.  The procedure are shown in Figure 1. 

In the first iteration, the procedure initially constructs single-gene classifiers and adds the gene that provides highest initial classification accuracy to the gene set. The algorithm, then, samples $c$ cells misclassified by the initial single-gene classifier out of the total set of $N$ cells and adds them to the cell set. The parameter $c$ is determined by the user according to the nature of dataset and available computational resources. The algorithm trains an SVM on the cell set using the current gene set, which defines an SVM margin $w$ that optimally separates cells into classes that are consistent with labels on the cell set. Using the SVM classification, the algorithm identifies cells that have been misclassified with the initial gene set. The algorithm, then, identifies genes that can be added to the gene set to improve performance on the misclassified examples.

To identify maximally informative genes,  we developed a gene selection strategy, Max Margin Rotation (MMR), that evaluates all candidate genes and selects the gene that induces maximum rotation of the margin $w$. The ActiveSVM algorithm continues iteration until a max gene number, $k$, is reached. The max gene number $k$ can be set as any integer smaller than $M$ and can be set to small values during exploratory analysis and to larger values for more exhaustive exploration of a data set. The integrated algorithm is shown in Algorithm 1. 

The most important feature of our ActiveSVM procedure is that the algorithm must never load an entire data set into memory. At each step, the procedure performs classification of cells using a minimal gene set, and then performs detailed (all genes) analysis of only a subset of misclassified cells. Due to the design of the procedure, ActiveSVM can analyze large data sets that do not easily fit in memory.  In conventional SVM based feature selection, the user would first train an SVM classifier on the complete data set and then select features according to the absolute values of the components of weight $w$.\cite{pmlr-v3-chang08a}. We note that conventional feature selection procedures typically apply classification accuracy for feature selection. Conventional SFS often selects features based upon improvement in classification accuracy. We found empirically that MMR provides improved classification results and so selected MMR as our gene selection strategy.  

Based on the above outline of ActiveSVM, we can formalize the specific gene and cell selection strategies into two defined rules. For notation, in single-cell gene expression data, we use $x_i^{(j)} \in \mathbb{R}$ to denote the measurement of the $j$-th gene of the $i$-th cell. We assume the classification labels are given and consider a data-set $\{ x_i, y_i\}_{i \in \{1,\dotsc,N\}}$ contains $N$ cells with total $M$ genes, where $x_i=[x_i^{(j)}]_{j \in \{1,\dotsc,M\}}$ and $y_i \in \mathbb{Z}^N$ are labels. The labels could be binary or multi-class and can be derived from clustering. We also denote the gene expression vector of $i$-th cell with part of genes as $x_i^{(D)}=[x_i^{(j)}]_{j \in D}$, where $D \subset \{1,\dotsc,M \}$. And we use $J$ and $I$ to refer to the set of selected genes and cell set.

We assume the SVM classifier notation of one observation is $h_{w,b}(x_i^{(D)})=g(w^{T}x_i^{(D)}+b)$ for any $i \in \{1,2,\dotsc,N\}$ and $D \subset \{1,2,\dotsc,M\}$ with respect to observation $x \in \mathbb{R}^{|D|} $, where $w \in \mathbb{R}^{|D|}$ and $b \in \mathbb{R}$ are parameters (the margin and bias respectively). Here, $g(z)=1$ if $z \geq 0$, and $g(z)=-1$ otherwise. And the loss function is Hinge Loss\cite{rosasco2004loss} $ \text{loss}_i=\text{max}\{0,1-y_i (w^T x_i^{(D)}+b)\}$, where $y_i \in \mathbb{R}$ is the ground truth label of observation $x_i$.

\subsection*{Cell selection: identification of maximally informative cells}

For the cell selection strategy, we simply choose cells with largest SVM classification loss. The purpose of cell selection is to use the most maximally informative cells as a smaller training set to select the next gene. In SVM classifier, samples separable in $n$-D are also separable in $(n+1)$-D as they are at least separated by the same boundary with zero at the $(n+1)$-th dimension. Therefore, to improve the accuracy with a new gene, we must only consider the misclassified cells. We identify such cells through analysis of the dual form of the classical SVM classification problem. After solving the primal optimization problem of soft margin SVM, we have the dual optimization problem with a non-negative Lagrange multiplier $\alpha_i \in \mathbb{R}$ for each inequality constraint.\cite{bottou2007support} 

\begin{equation}
\begin{aligned}
\max_{\alpha} \quad & \sum_{i=1}^{N}{\alpha_{i}}-\frac{1}{2}\sum_{i_1,i_2=1}^{N}{y_{i_1}y_{i_2}\alpha_{i_1}\alpha_{i_2}<x_{i_1}^{(J)},x_{i_2}^{(J)}>}\\
\textrm{s.t.} \quad & 0\leq \alpha_{i}\leq C\\
  &\sum_{i=1}^{N}{\alpha_{i}y_{i}}=0    \\
\end{aligned}
\end{equation}

Here $x_{i}^{(J)}$ refers to the measurement of the $i$-th cell with all selected genes, and $C \in \mathbb{R}$ is a hyper-parameter we set to control the trade-offs between size of margin and margin violations when samples are non-separable.

We solve the optimal solution $\alpha^*$ and apply the Karush-Kuhn-Tucker(KKT) dual-complementarity conditions\cite{gordon2012karush} to obtain the following results where $w \in \mathbb{R}^{|J|}$ and the intercept term $b \in \mathbb{R}$ are optimal.

\begin{equation}
\begin{aligned}
    \alpha^*_{i}=0 &&&\Rightarrow&& y_{i}(w^{T}x_{i}^{(J)}+b)> 1\\
    \alpha^*_{i}=C &&&\Rightarrow&& y_{i}(w^{T}x_{i}^{(J)}+b)< 1\\
    0<\alpha^*_{i}<C &&&\Rightarrow&& y_{i}(w^{T}x_{i}^{(J)}+b)= 1.\\
\end{aligned}
\end{equation}

Therefore, for each cell, the Lagrange multiplier $\alpha_i$ indicates whether the cell falls within the SVM margin defined by the vector $w$. $\alpha_{i}>0$ means $y_{i}(w^{T}x_{i}+b)\leq 1$, i.e. cells are on or inside the SVM margin. Hence, we can directly select cells with $\alpha_{i}>0$.  In practice, we normally only select cells with $\alpha_{i}=C$, which indicates incorrectly classified cells.

\subsubsection*{Discussion of min-cell and min-complexity cell sampling strategies}

Using this mathematical formulation, we develop two different versions of the ActiveSVM procedure, the min-complexity strategy and min-cell strategy, for distinct goals. The min-complexity strategy minimizes the time and memory consumption when computational resources are restricted or where a user desires to reduce run-time. 
In the min-complexity strategy, a certain fixed number of cells is sampled among all misclassified cells and used as the cell set for gene selection in each iteration. Therefore, a small number of cells can be analyzed at each round and typically only few cells might be selected repeatedly. 

In the min-cell strategy, to reduce the number of unique cells required, the misclassified cells already used in previous steps are given the highest priority to select again. Therefore, the min-cell strategy attempts to re-use cells across rounds of iteration and  aims to minimize the total number of unique cells we acquire during the entire procedure. The min-cell strategy can be applied to limit the number of cells required to perform the analysis in settings where cell acquisition might be limiting including in the analysis of rare cell populations or in clinical data sets. 

For the min-cell strategy, assume we select $c$ cells for each iteration and there are $a+b$ misclassified cells at the current iteration, where $a$ cells have been used at least once in previous iterations while $b$ cells are new cells. If $a \geq c$, we do not need to add any new cells to current cell set. If $a<c$, we sample $c-a$ cells among the $b$ new cells. Then the algorithm uses the whole cell set for the next gene selection step. When using the min-cell strategy, cells tend to be re-used many times and the curve of number of unique cells we acquire converges to a fixed value along with the number of genes we select. In experiments, the number of cells selected for each step, $c$, is a hyper-parameter set by the user. Typically, the parameter can be set to a small number using the min-complexity strategy, as a sufficient number of new cells is considered in the procedure. Selecting a small number of cells each round reduces computational complexity. In the min-cell strategy it can be advantageous to select a larger number of total cells to guarantee diversity of training cells while still bounding the total number of cells used.

\subsubsection*{Balancing cell-sampling across cell-classes}
In addition to the min-cell and min-complexity options, we also include two version of cell sampling strategies. The first one is uniform, random sampling. Another option is cell `balanced' sampling that can be applied to balance sampling across a series of cell classes. In the `balanced' strategy, we sample a fixed number of cells from each cell class, and for classes with insufficient cells we sample all the cells in the class. Mathematically, assume there are $Z$ classes and $S$ is the set of all misclassified cells this step. We should sample $c'$ cells from a candidate cell set, $S'$, for the current iteration. In min-complexity strategy, $c'=c$ and the candidate cell set, $S'$, should be $S$ itself. For the min-cell strategy, $c'=c-\min\{c,|I\cap S|\}$, where $I$ is the cell set before current iteration, and the candidate cell set $S'=S\setminus I$. Assume $S'= \cup_{z=1}^Z S'_z$, where $S'_z$ are the set of cells in class $z$, and $|S'_z| \le |S'_{(z+1)}|$ for any $z \in \{1,2,...,Z-1\}$. We sample cells in order from class $1$ to class $Z$ and denote $P_z$ as the union set of all selected cells from all classes after class $z$. Then, for class $z$, if $|S'_z| \le (|S'|-|P_{z-1}|)/(Z-z+1)$, we select all cells in $S'_z$. Otherwise, if $|S'_z| > (|S'|-|P_{z-1}|)/(Z-z+1)$, we randomly sample $(|S'|-|P_{z-1}|)/(Z-z+1)$ cells in $S'_z$. The procedure repeats for all classes and then we have $P_Z$ as the cells we select at this iteration.

\subsection*{Gene selection by maximizing margin rotation}

To select maximally informative genes at each round, we analyze misclassified cells and identify genes that will induce the largest rotation of the classification margin. Our procedure is inspired by the active learning method, Expected Model Change\cite{settles2009active}. We quantify rotation of the margin by calculating the twist angle induced in $w$ when we add a new dimension (gene) to the classifier. Assume $J$ is the set of genes we have selected so far. Once we add a gene into the $|J|$-dimensional data space, the parameter $w$ will have one more dimension. The rotation of margin measures how much $w$ twists after adding the new dimension compared with weight in the previous iteration.

Specifically, assume $J$ is the set of genes we have selected so far. We derive the corresponding $w$ from the optimal solution $\alpha^*$.\cite{bottou2007support} After solving the dual optimization problem (1), we have:

\begin{equation}
     w = \sum_{i\in I}{\alpha^*_{i}y_{i}x_{i}^{(J)}}.
\end{equation}

Then we pad $w$ with zero to get a $|J+1|$-dimensional weight $w_{padded}$, whose first $|J|$ dimensions is $w$ and the $|J+1|$-th dimension is zero.

For each candidate gene $j$, we train a new $|J+1|$-dimensional SVM model and have weight $w_j$,where $j \in \{1,\dotsc,M\} \setminus J$. That is to say, for candidate gene $j$, we solve the dual optimization problem (4) and find a new optimal multiplier $\alpha^{*(j)}$. Note that we only use the selected cells here, $i_1,i_2 \in I$. 

\begin{equation}
\begin{aligned}
\max_{\alpha} \quad & \sum_{i \in I}{\alpha_{i}^{(j)}}-\frac{1}{2}\sum_{i_1,i_2 \in I}{y_{i_1}y_{i_2}\alpha_{i_1}^{(j)}\alpha_{i_2}^{(j)}\langle x_{i_1}^{(J \cup \{j\})},x_{i_2}^{(J\cup \{j\})} \rangle }\\
\textrm{s.t.} \quad & 0\leq \alpha_{i}^{(j)}\leq C\\
  &\sum_{i \in I}{\alpha_{i}^{(j)}y_{i}}=0    \\
\end{aligned}
\end{equation}

Then we have $w_j$ as shown in equation (5):

\begin{equation}
     w_j = \sum_{i \in I}{\alpha_{i}^{*(j)}y_{i}x_{i}^{(J \cup \{j\})}}
\end{equation}

The angle $\theta_j$ between $w_j$ and $w_{padded}$ is the expected angle the margin rotates, corresponding to the $j$-th candidate gene. Then the $j$-th gene with largest angle $\theta_j$ will be selected. We measure the angle between two vectors using cosine similarity\cite{xia2015learning}:

\begin{equation}
\begin{aligned}
    \vartheta_j=\arccos{\cos{\vartheta_j}} = \arccos{ \frac{ \langle w_j,w_{padded} \rangle}{\parallel w_j \parallel \parallel w_{padded} \parallel}}
\end{aligned}
\end{equation}

Therefore, a new gene, which maximizes $\vartheta_j$, is selected to maximize the expected model change.

For multi-class classification, the SVM is handled according to a one-vs-rest scheme\cite{bishop2006pattern}, where a separate classifier is fit for each class, against all other classes. Margin rotation is represented as the sum of weight components in each class dimension. Hence with $Z$ classes, we get $Z$ weight components corresponding to $Z$ one-vs-the-rest classification decision boundaries. Assume the weight component for class $z$ of the previous $|J|$-dimensional SVM model is $w^{(z)}$. Denote the $|J+1|$-dimensional weight after zero-padding of $w^{(z)}$ as $w^{(z)}_{padded}$ and the new $|J+1|$-dimensional weight component of class $z$ with $j$-th gene as $w_j^{(z)}$, where $z \in {1,\dotsc,Z}$. Then we have:

\begin{equation}
\begin{aligned}
    \vartheta_j^{(z)}=\arccos{\cos{\vartheta_j^{(z)}}} = \arccos{ \frac{\langle w_j^{(z)},w_{padded}^{(z)} \rangle }{\parallel w_j^{(z)} \parallel \parallel w_{padded}^{(z)} \parallel}}
\end{aligned}
\end{equation}
\begin{equation}
\begin{aligned}
    \vartheta_j=\sum_{z=1}^Z \vartheta_j^{(z)}
\end{aligned}
\end{equation}

\begingroup
\LinesNumberedHidden
\begin{algorithm}[H]
\SetAlgoLined
\DontPrintSemicolon 
    \KwIn{$c,k \in \mathbb{N}$, $J= \emptyset$}
    \KwOut{$J$}
    Randomly or `balanced' select $c$ cells $I \subset \{1,\dotsc,N\}, |I|=c$
    
    Train a 1-D SVM model on training set $I$ for each candidate gene: $\{h^{(j)}_{w,b}\}_{j \in \{1,\dotsc,M\}}$
    
    $loss_j= \sum _{i \in I} \max\{0,1-y_ih^{(j)}_{w,b}(x_i^{(j)})\}$
    
    Select one gene $j_0 \in \{1,\dotsc,M\}$ with lowest $loss_j$

    $J= J \cup \{j_0\}$
    
  \Repeat{$|J| \geq k$}{
    Optimize (1) and get optimal solution $\{\alpha_i^* \}_{i=1}^N$
    
    Get the the set of misclassified cells $S \subset \{1,\dotsc,N \}$ with $\alpha_i^*=C$
    
    \eIf{min-complexity}
    {
        Randomly or `balanced' select $c$ cells $I \subset S$, where $|I|=c$;
    }{
    \If{min-cell}
    {
        $c'=\min\{c,|I \cap S|\}$;

        Randomly  or `balanced' select $c-c'$ cells $I' \subset S\setminus I$, where $|I'|=c-c'$;
        
        $I= I \cup I'$
    }
    }
    
    $w = \sum_{i\subset I}{\alpha^*_{i}y_{i}x_{i}^{(J)}}$
    
    $w_{padded}=[w,0]$
    
    For each $j \in \{1,\dotsc,M\} \setminus J$, optimize (4) and get optimal solution $\{\alpha_i^{*(j)} \}_{i \in I}$
    
    $w_j = \sum_{i \in I}{\alpha_{i}^{*(j)}y_{i}x_{i}^{(J \cup \{j\})}}$
    
    $\vartheta_j=\arccos{\cos{\vartheta_j}} = \arccos{ \frac{<w_j,w_{padded}>}{\parallel w_j \parallel \parallel w_{padded} \parallel}}$
    
    Select one gene $j^* \in \{1,\dotsc,M\} \setminus J$ with largest $\vartheta_j$
    
    $J= J \cup \{j^*\}$

    }
 
  \caption{Active Linear SVM Gene Selection}
  \label{algo:b}
\end{algorithm}
\endgroup

\subsection*{Memory complexity}
One of the key contribution of ActiveSVM is that it significantly saves memory usage because only a small part of data is used at each iteration. The entire dataset can be stored in disk and the algorithm only loads two small matrices into memory, a $N \times |J|$ matrix of all cells with the currently selected genes and a $|I| \times M$ matrix of the cell set with all genes. The memory complexity is $\mathcal{O}(M+N)$ while the memory complexity of algorithms using the entire dataset should be at least $\mathcal{O}(MN)$.
The min-cell strategy minimizes the total number of unique cells acquired to reduce the cost of data measurement, acquisition and storage. 

\subsection*{Time complexity}
The time complexity of the complete procedure depends primarily on the training of SVM. The standard time complexity of SVM training is usually $\mathcal{O}(M N^2)$ \cite{abdiansah2015time,scikitSVM}. Assume that we plan to select $k \in \mathbb{N}$ genes in total and use the cell set $I_i$ of poorly classified cells at $i$-th iteration, where $k, k^2 \ll M$ and $|I_i|, |I_i|^2 \ll N$ are constants. Then the computational complexity of ActiveSVM is:

\begin{equation*}
    \begin{aligned}
    \mathcal{O}(\sum_{i=1}^k{(i \cdot N^2+(M-i) \cdot (i+1) \cdot |I_i|^2))}\sim \mathcal{O}(N^2+M).
    \end{aligned}
\end{equation*}

The key reduction in total complexity occurs because each step is performed using $N$ cells with of order $k, k^2 \ll M$ genes or using order $M$ genes with $|I_i|$ cells. Therefore, the polynomial $\mathcal{O}(M N^2)$ is reduced to two separate steps that are individually $\mathcal{O}(N^2)$ and $\mathcal{O}(M)$. 

And in practice, we implement ActiveSVM using the linear SVM library LIBLINEAR\cite{fan2008liblinear}, whose time complexity is $\mathcal{O}(MN)$. Therefore, 

\begin{equation*}
    \begin{aligned}
    \mathcal{O}(\sum_{i=1}^k{(i \cdot N+(M-i) \cdot (i+1) \cdot |I_i|))}\sim \mathcal{O}(N+M), 
    \end{aligned}
\end{equation*}

and the corresponding time complexity of ActiveSVM with LIBLINEAR is $\mathcal{O}(M+N)$. 

In the gene selection part, the margin rotation angles of all candidate genes can be computed in parallel, which also accelerates the algorithm.
The complexity provides a significant improvement in marker gene selection methods especially for large-scale datasets.

\subsection*{ActiveSVM can incorporate cell labels derived from unsupervised analysis, experimental conditions, or biological knowledge} 
The goal of ActiveSVM is to discover minimal gene sets for extracting biological information from single-cell data sets. To define minimal gene sets, we apply a classification task in which we find genes that enable a SVM classifier to distinguish single-cells with different labels ($y_i$). In practice, explicit cell-type labels are often not known for a data set. An extremely common work-flow in single-cell genomics applies Louvain clustering algorithms to identify cell classes and visualizes these cell classes in UMAP or tSNE plots (\cite{macosko2015highly,wolf2018scanpy}. The cell clusters output by clustering work-flows in commonly used single-cell analysis frameworks provide a natural set of labels for down-stream analysis. In fact, ActiveSVM can, then, identify specific marker genes for interpreting the identified cell-clusters and determining their biological identify. More broadly, cell-class labels can be quite general including the identity of a genetic perturbation (Figure 6), the spatial location of a cell (Figure 7). We can imagine the application of ActiveSVM to a broad set of additional labels including membership to a differentiation trajectory or lineage tree \cite{street2018slingshot}.

\section*{Results}

We test our ActiveSVM feature-selection method on four single-cell mRNA-seq datasets: a dataset of peripheral blood mononuclear cells (PBMCs)\cite{zheng2017massively}, the megacell 1.3 million cell mouse brain data set \cite{genomics20171}, the Tabula Muris mouse tissue survery dataset\cite{tabula2018single}, and a multiple myeloma human disease dataset \cite{chen2020dissecting}. Later, we demonstrate generalization of the strategy to additional types of single-cell data analysis, including a perturb-seq dataset where genes impacted by Cas9 based genetic perturbation, and a spatial transcriptomics dataset by seqFish+.

For each analysis, we show the classification accuracy of the test set along with the number of genes we select. We also compare the classification performance to several widely-used feature selection methods, including conventional SVM, correlation coefficient\cite{taylor1990interpretation}, mutual information\cite{vergara2014review}, Chi-square\cite{mchugh2013chi}, feature importance by decision tree\cite{safavian1991survey}, and randomly sample genes, showing that ActiveSVM obtains the highest accuracy. All of the comparison methods select genes one by one and select a new gene with the largest score in terms of the corresponding evaluation functions while using the same number of cells as our method. However, all methods randomly sample cells at each iteration without an active learning approache. For perturb-seq and seqFish+ datasets, we also show the accuracy performance of comparison methods, where the entire dataset is used. Specifically, conventional SVM based feature selection also called naive SVM selects the gene with largest weight component, which is the most popular SVM feature selection method. In our application of ActiveSVM, we tested both the min-cell strategy and min-complexity strategies as well as randomly sampling and 'balanced' sampling. 

In each experiment, the data set was first pre-processed and normalized using standard single-cell genomics strategies (See Data Pre-processing). The entire dataset was, then, randomly split into training set with the size of $80\%$ and test set with the size of $20\%$. For conventional and ActiveSVM, we found the approximately optimal parameter by grid-search \cite{syarif2016svm} across lists of candidate values for some key parameters in the framework of 3-fold cross validation \cite{arlot2010survey}. The optimal parameters were fixed during all iterations. For the comparison methods, we use 3-fold cross validation grid-search to obtain the optimal parameters at each single iteration. We also implemented the algorithms called $\texttt{min\_complexity\_cv}$ and $\texttt{min\_acquisition\_cv}$ that apply grid-search and cross validation for each single SVM trained in each iteration (see  Code Availability). The parameter setting details are shown in Parameters section. 

In our evaluation, besides accuracy curves with proportion confidence interval\cite{brown2001interval}, we also show the distribution of gene markers we selected and the relation with classification target. The subplots include the gene expression values on t-SNE projection, the mean of each class, histogram distribution, violin plot, the correlation coefficient heatmap, etc. 

To indicate the efficiency, we also recorded the run time, peak memory usage, and the total number of unique cells we used of ActiveSVM on these datasets. We used r5n.24xlarge\cite{r5n}, a type of EC2 \cite{ec2} virtual server instance on AWS\cite{aws}, with $96$ virtual central processing units (vCPU) and $768$ GiB memory on Linux\cite{linux} system. For example, we selected 50 genes on the largest dataset, mouse brain `megacell' dataset, which contains $1306127$ cells and $27998$ genes, using ActiveSVM and some other popular feature selected methods, including mutual information, feature importance by decision tree, and conventional SVM. The peak memory usage of ActiveSVM is 2111 MB while other methods all consume more than 78600 MB. The run time of the min-complexity method is about 69 minutes and of the min-cell method is about 243 minutes. Each comparison method takes more than 4 days on the same server machine. The run time and peak memory usage of ActiveSVM on all six datasets are shown in Table 1. The ActiveSVM package used for the brain megacell dataset only loads the selected genes and cells into memory at each iteration while other two experiments called the package loading the entire dataset. Both packages are provided in Code Availability Section. 

\begin{table}[!htbp]
\centering
\caption{Run Time and Peak Memory Usage of ActiveSVM.}
\begin{tabular}{*6c}
\toprule
{} & matrix size & min-complexity & min-cell run& memory & unique cells\\
{} &(cells, genes) & run time (s/gene) & time (s/gene) &  (MB) & (min-cell)\\
\midrule
mouse megacell  & (1306127, 27998) & 4142/50 & 14580/50 & 2111 & 712\\
PBMC & (10194, 6915) & 121/50 & 176/20 & 1325 & 298\\
Tabula Muris & (55656, 8661) & 737/150 & 7701/100 & 1093 & 779\\
MM & (35159, 32527) & 127/40 & 449/40 & 1616 & 445\\
seqFish & (913, 10000) & 33/30 & 728/30 & 887 & 428\\
perturb-seq & (10895, 15976) & 3424/50 & {} & 9493 & 3827\\
\bottomrule
\end{tabular}
\end{table}

\subsection*{Active feature selection on human PBMC data}

\begin{figure}
    \centering
    \includegraphics[width=\textwidth]{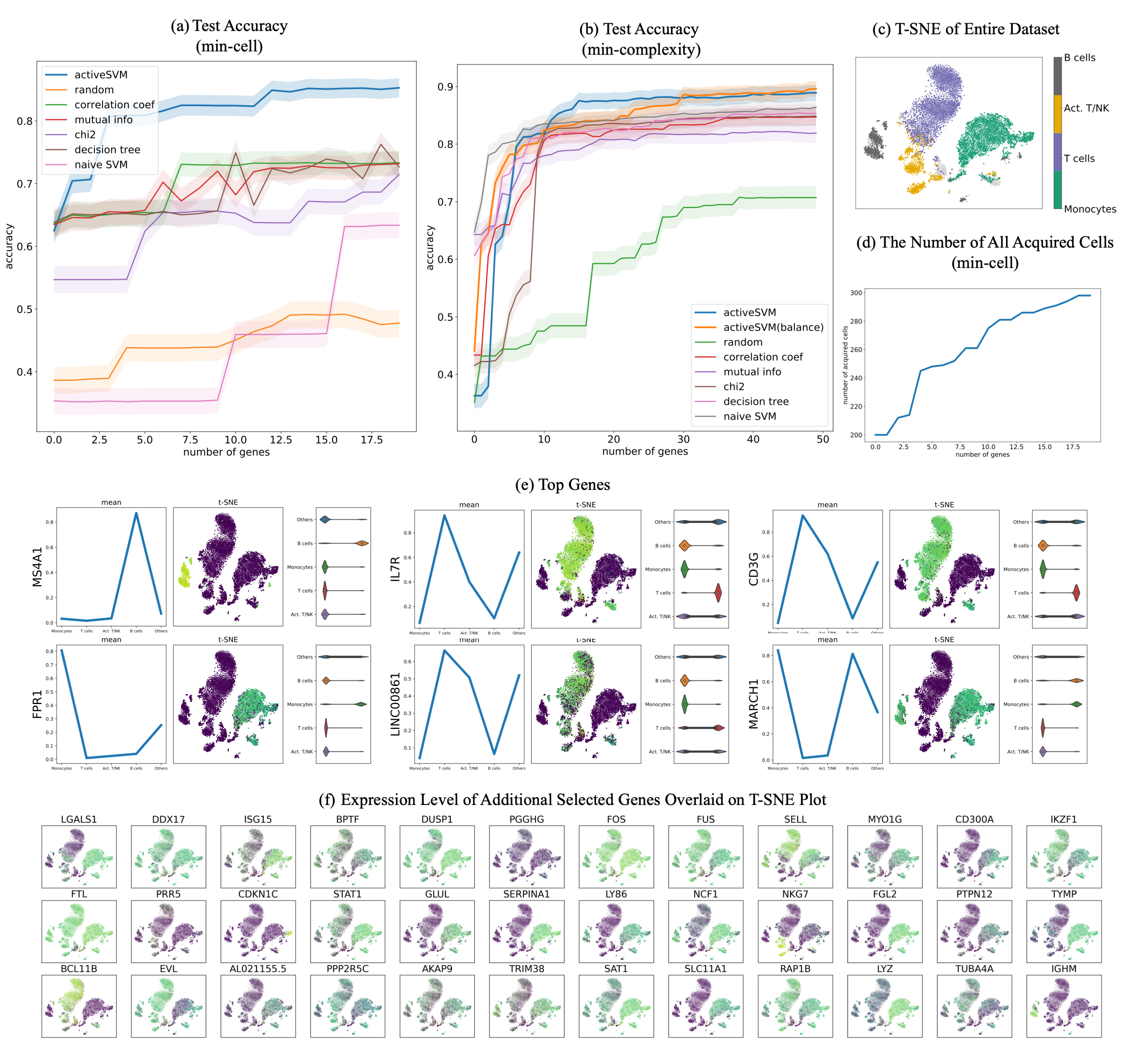}
    \caption{\textbf{Gene selection and cell-type classification for PBMC dataset.} (a) The test accuracy for min-cell strategy and a series of comparison classification strategies. The min-cell strategy selects $k=20$ genes and select $c =100$ cells each iteration with confidence interval estimates; (b) The test accuracy of min-complexity strategy that selects $k=50$ genes using $c=20$ cells each iteration; (c) The t-SNE plots of the entire filtered dataset; (d) The total number of unique cells used vs gene set size with the min-cell strategy; (e) plots showing the expression of several genes markers, including mean on classes, gene expression value on t-SNE projection, and violin plots; (f) expression level of additional selected genes overlaid on t-SNE plot.}
    \label{fig:PBMC}
\end{figure}

To test the performance of ActiveSVM, we used the method to extract classifying gene subsets for human PBMCs. We analyzed a single-cell transcriptional profiling data set for 10194 cells \cite{zheng2017massively} with 6915 genes. We used Louvain clustering \cite{blondel2008fast} to identify T-cells, activated T/NK cells, B-cells, and  Monocytes (Figure 2(c)). 


The min-cell strategy classified the 5 major cell-types at greater than $85 \%$ accuracy with as few as $15$ total genes (Figure 2(a)) and the test accuracy of min-cell, with both randomly sampling and 'balanced' sampling, also reached much higher accuracy than the comparison methods. 

A key benefit of the active learning strategy is that a relatively small fraction of the data set is analyzed, so that the procedure can generate the gene sets while only analyzing 298 cells (Figure 2(d)). At each iteration, a specific number of misclassified cells ($c=100$) are selected but the total number of cells used does not increase in increments of $100$, since some cells are repeatedly misclassified and are thus repeatedly used for each iteration.

In addition to enabling cell-type classification of the data set, the ActiveSVM gene sets provide a low-dimensional space in which to analyze the data. When we reduced our analysis to consider only the top 100 genes selected by the ActiveSVM algorithm, we were able to generate a low-dimensional representations of the cell population (t-SNE) that preserved critical structural features of the data, including the distinct cell-type clusters (Figure 2(c)).

The procedure generates gene sets that contain known and novel markers, each plotted individually in a t-SNE grid (Figure 2(e)(f)). For instance, MS4A1 and CD79 are well-established B-cell markers, and IL7R and CD3G are well-established T-cell markers. However, we also find genes which are not commonly used as markers, but whose expression is cell-type specific. For instance, we find highly monocyte-specific expression of FPR1, which encodes N-formylpeptide receptor, which was recently discovered to be the receptor for plague effector proteins \cite{Osei-Owusu2019-zl}. We also find T-cell/NK-cell specific expression of a long noncoding RNA, LINC00861, whose function is unknown but has been correlated with better patient outcome in lung adenocarcinoma \cite{Sage2020-et}. The marker genes are generally highly specific for individual cell types, but some mark multiple cell types (i.e. MARCH1, which marks monocytes and B-cells).

\subsection*{Scaling of ActiveSVM feature selection to million cell dataset}

\begin{figure}
    \centering
    \includegraphics[width= .9 \textwidth]{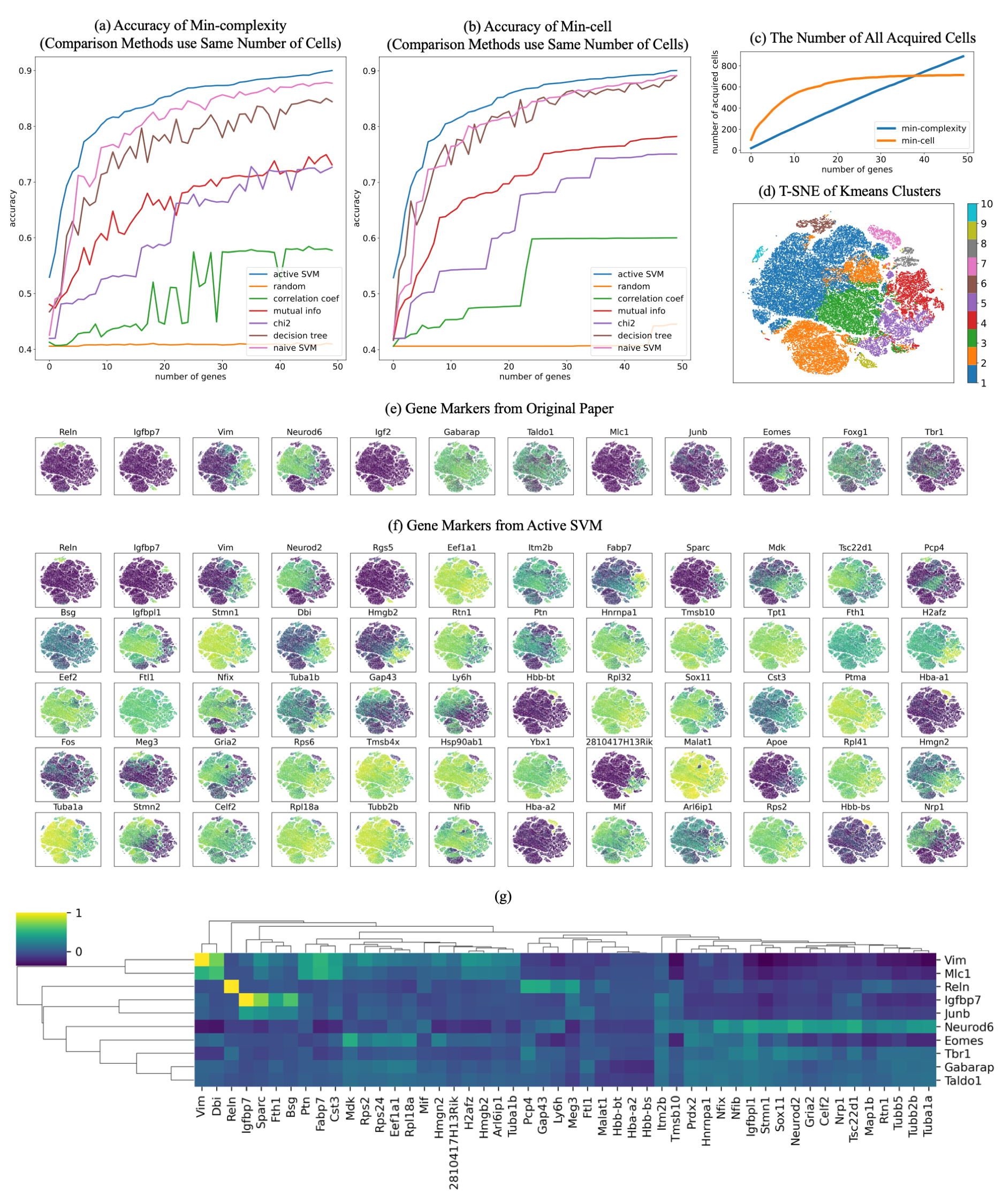}
    \caption{\textbf{Scaling of ActiveSVM feature selection to 1.3 million cell mouse brain data set} (a) The test accuracy of min-complexity strategy that selects 50 genes using 20 cells each iteration; (b) The test accuracy of min-cell strategy that selects 50 genes using 100 cells each iteration; (c) The total number of unique cells used vs gene set size with both the min-complexity and the min-cell strategy; (d) The t-SNE plots of the entire filtered dataset with 10 classes by k-means clustering; (e) expression level of the gene markers from previously published analysis overlaid on t-SNE plot; (f) expression level of the gene markers selected by ActiveSVM overlaid on t-SNE plot, where the first row are the genes that have similar distribution with gene markers from previously analysis and other genes are new markers correlated with the classification target. (g) Correlation matrix of literature markers (y-axis) from \cite{bhaduri2018identification} versus ActiveSVM selected genes (x-axis).} 
    \label{fig:brain}
\end{figure}

To demonstrate the scaling of the ActiveSVM feature selection method to large single cell mRNA-seq data sets, we applied the method to extract compact gene sets from the 10x genomics the `megacell' demonstration data set \cite{genomics20171}. The megacell dataset was collected by 10x genomics as a scaling demonstration of their droplet scRNA-seq technology. The data set contains full transcriptome mRNA-seq data for $1.3$ million cells from the developing mouse brain profiled at embryonic day 18 (E18) \cite{genomics20171}. The data set is one of the largest single cell mRNA-seq data sets currently available. The size of the data set has been a challenge for data analysis, and a previous analysis paper was published that developed sub-sampling methods that extract marker genes and cell-types by extracting sub-sets of of the data set containing $\sim 100,000$ cells \cite{bhaduri2018identification}. 
We applied our ActiveSVM method to extract minimal genes sets for classifying the $10$ classes of cells that were extracted through k-means\cite{likas2003global} clustering in the internal analysis of the data (Figure 3(a)(b)). The min-complexity algorithm used $20$ cells at each iteration and the min-cell algorithm selected $100$ cells each loop. The min-cell algorithm acquired fewer unique cells, as cells are selected repeatedly (Figure 3(c)). On this dataset, both algorithms use 'balanced' sampling for both min-complexity and min-cell strategies. As the dataset is too large to produce t-SNE, we randomly sampled $30,000$ cells and find the tSNE projection, which is shown with the input cell clusters in Figure 3(d). 

While the size of the data set has presented challenges for conventional sampling methods, the ActiveSVM algorithm must only acquire from memory a small number of genes or cells at each round of analysis, and therefore, the method avoids computing across the entire 1.3 million cells and $\sim 30,000$ gene data set. We found that it was possible to run ActiveSVM on a conventional lap-top. For decreasing compute time, we analyzed the megacell data set on an AWS instance r5n.24xlarge. On this instance, ActiveSVM ran in 69 minutes for the min-complexity strategy and 243 minutes for the min cell strategy. As a comparison, naive SVM required greater than four days of computation to run on all 1.3 million cells on the same AWS instance (Table 1). 

To provide a bench-marking for ActiveSVM, we instead compared the accuracy of ActiveSVM to a data set where we allow ActiveSVM to run on the data set;we extract the number of analyzed cells, and then provide this same number of cells to the other methods shown in figure 3(a)(b). Applying the other methods to sub-sampled data, allowed us to extract the classification accuracy as a bench-marking for ActiveSVM. 

In addition to performing the classification task, the ActiveSVM procedure discovers gene sets that achieve $\sim 90\%$ classification accuracy with only 50 genes. The procedure discovered a series of cluster specific marker genes that extend prior analysis. For example, the analysis in \cite{bhaduri2018identification} identified marker genes through sub-sampling and prior biological literature. A set of genes identified previously is shown in Figure 3(e). The ActiveSVM analysis discovered several of the same markers as the previous work (Reln, Vim, Igfbp7) (Figure 3(f)).

Further, ActiveSVM extended previous analysis by identifying additional markers that correlate with the previously analysis as well as marker genes of additional cell states. The development of radial glial cells, in particular, has been of intense recent interest because radial glial cells are the stem cells of the neocortex in mouse and human \cite{pollen2015molecular}. Careful molecular analysis has defined markers of radial glial cells including Vim. ActiveSVM identified a group of genes whose expression correlates with Vim across the E18 mouse brain. Our analysis identified an additional set of genes expressed in the same cell population as Vim including, Dbi (Diazepam Binding Inhibitor, Acyl-CoA Binding Protein), Hmgb2, and Ptn. A correlation matrix (Figure 3g) showing the correlation of ActiveSVM identified genes (x-axis) with literature markers (y-axis) discussed in \cite{bhaduri2018identification} reveals the existence of Vim correlated genes.  The Vim genes were of interest because they include additional transcription factors Hmgb2 \cite{pollen2015molecular} and also a core group of genes, Ptn and Fabp7 (also Brain Lipid Binding Protein), two components of a radial glia signaling network  \cite{anthony2005brain,andrews2020mtor,pollen2015molecular} that has been identified as a core regulatory module supporting the proliferation and stem cell state in the radial glial cell population.  

The neural progenitor transcription factor Neurod6 marked a separate cell population that we identified to contain genes including Neurod2 (a transcription factor) and Sox11 (a transcription factor) as well as glial transcription factors Nfib and Nfix and the receptor Gria2 (Glutamate Ionotropic Receptor AMPA Type Subunit 2). The marker genes observed in Neurod6 expressing cells were anti-correlated with the Vim correlated markers suggesting that ActiveSVM identified two distinct regulatory modules. Structurally, the tubulin proteins Tuba1b and Tuba1a were expressed in Vim and Neurod6 populations respectively. In addition to genes correlated or anti-correlated with existing markers, ActiveSVM identified markers of additional cell populations including Meg3, a long non-coding RNA expressed in cluster 2. 

Broadly, the analysis of the `megacell' mouse brain data set demonstrates that ActiveSVM scales to analyze a large data set with $>1$ million cells. The analysis of such large data sets has been challenging with conventional approaches that attempt to store the entire set in memory for analysis. Previous analysis of the 10x megacell dataset found that sub-samples with greater than $100,000$ cells would yield an out of memory error on a server node with 64 cores, a 2.6 GHz processor, and 512 GB of RAM \cite{bhaduri2018identification}. 
ActiveSVM iterates through analysis of cells and genes while focusing computational resources on poorly classified cells, and so ActiveSVM does not load the entire dataset into memory but can read cells and genes from disk as needed. Further, through iterative analysis, ActiveSVM identifies known marker and regulatory genes, genes that correlate with known markers as well as marker genes of additional cell populations that could provide a starting point for future experimental investigations.

\subsection*{Identifying gene sets for cell-type classification in the Tabula Muris tissue survey}

\begin{figure}
    \centering
    \includegraphics[width=\textwidth]{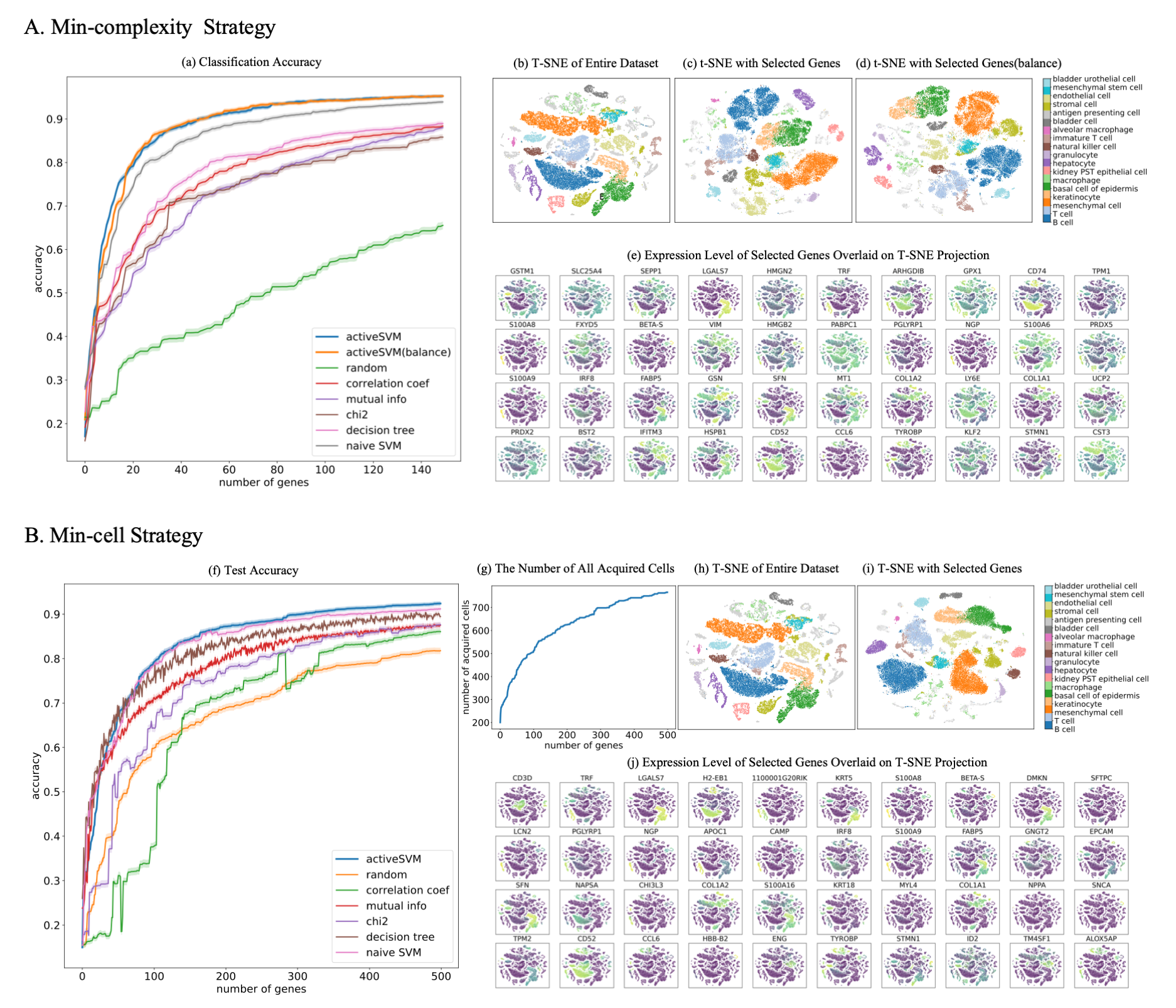}
    \caption{\textbf{Minimal gene sets for cell-type classification in the Tabula Muris mouse tissue survey} (A) Classification results of 150 genes selected using the min-complexity strategy with 20 cells each iteration. (B) 500 genes selected using the min-cell strategy with 200 cells per iteration. Results for standard and balanced strategy shown with comparison methods and confidence intervals. The subplots contain: classification accuracy vs gene set size using the min-complexity strategy (a) and min-cell strategy (f); the t-SNE plots of the entire filtered dataset (b)(h); the t-SNE plots of the gene set selected using min-complexity strategy with randomly sampling (c) and 'balanced' sampling (d), and gene set selected using the min-cell strategy (i); the expression level overlaid on t-SNE projection for genes selected by min-complexity (e) and by min-cell (j); and the total number of unique cells used vs gene set size with the min-cell strategy (g).}
    \label{fig:tabula}
\end{figure}

In addition to analyzing a data set with a large number of total cells, we sought to benchmark performance of ActiveSVM feature selection on a data set with a large number of distinct cell types. We applied ActiveSVM to the Tabula Muris mouse tissue survey, a droplet-based scRNA-sequencing data-set, that contains 55,656 single cells across 58 annotated cell types, and 12 major tissues \cite{tabula2018single}. For each cell, 8,661 genes are measured. In our analysis, we used the supplied cell-type labels, agnostic of tissue type. Thus, cells labeled `macrophage' from the spleen are considered to belong to the same class as cells labeled `macrophage' from the mammary gland.

Even with a large number of cell types, ActiveSVM can construct gene sets that achieve high accuracy ($>90\%$), compared to other methods (Figure 4(a)(f)). To construct a gene set of size 500, ActiveSVM feature selection used fewer than 800 unique cells (Figure 4(g)) or an average of $14$ cells per cell type. We were able to recreate the clustering patterns from the original data (Figure 4(b)(h) when analyzing the cells within the low dimensional t-SNE space spanned by the selected 150 genes (Figure 4(c)(d)) or 500 genes (Figure 4(i)).

Our approach allowed us to construct a set of marker genes able to identify mouse cell types across disparate tissues. Even when analyzing a large number of cell types, we were able to identify highly cell-type specific genes, such as CD3D, a well-established T-cell marker, or TRF (transferrin), which is selectively secreted by hepatocytes\cite{Guan2020-kf}, or LGALS7 (galectin-7), which is specific for basal and differentiated cells of stratified epithelium \cite{Magnaldo1998-ho}. However, given the functional overlap between different cell types, the genes within our set include many that mark multiple cell types. For instance, H2-EB1\cite{Stables2011-wh}, a protein important in antigen presentation, is expressed in B-cells and Macrophages, both of which are professional antigen presenting cells (APCs). Our analysis also identified cell type-specific expression for a number of poorly studied genes, such as granulocyte- and hepatocyte- specific expression of 1100001G20RIK (also known as Wdnm-like adipokine), which has previously only  been associated with adipocytes \cite{Wu2008-pj}. 

\subsection*{Extraction of gene sets for classification of disease state in peripheral blood cells from multiple myeloma patient samples}

\begin{figure}
    \centering
    \includegraphics[width=\textwidth]{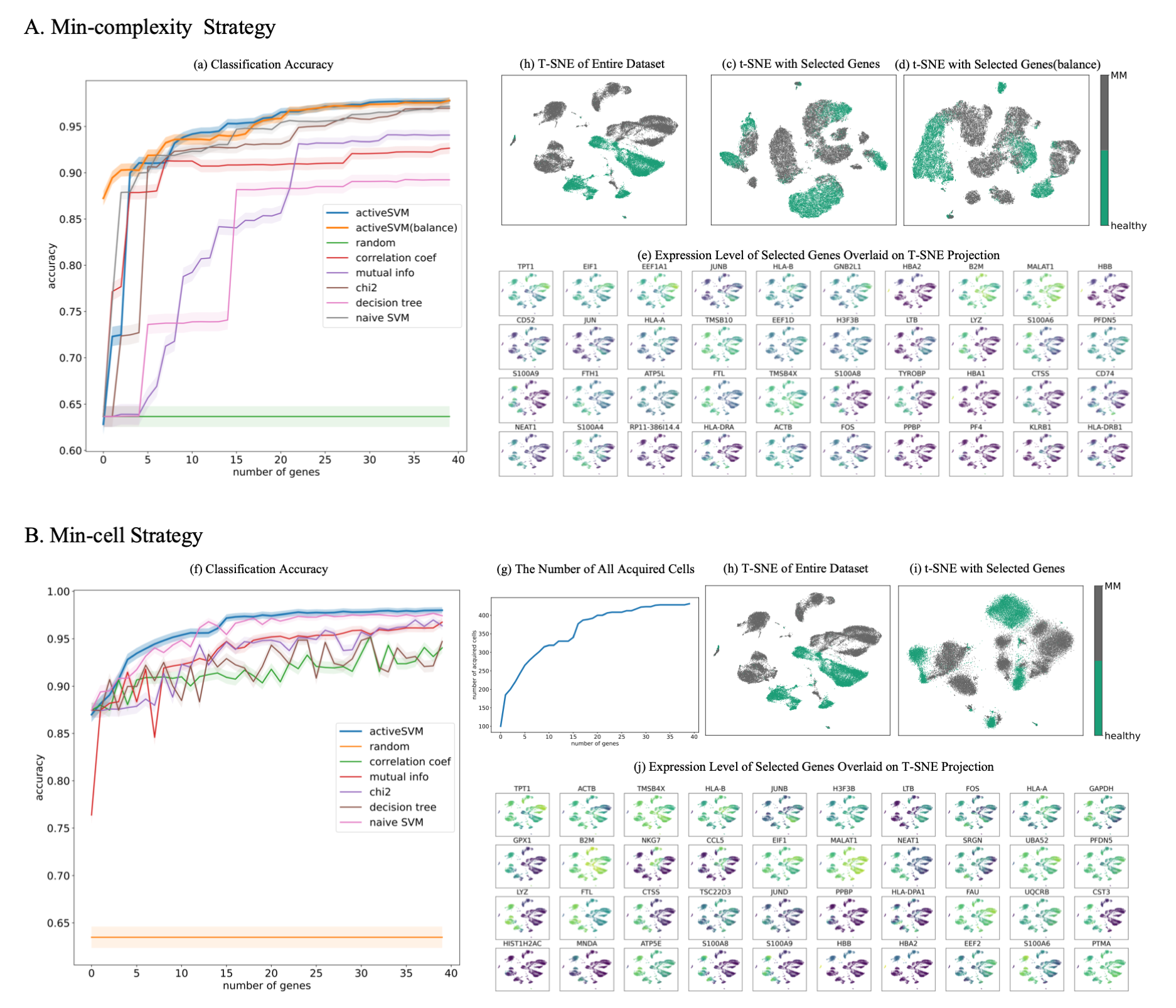}
    \caption{\textbf{Gene set selection for healthy vs disease classification in multiple myeloma dataset.} (A) classification results of 40 genes selected by min-complexity strategy using 20 cells each iteration. (B) 40 genes selected using Min-cell strategy with 100 cells per iteration. Results for standard and balanced strategy shown with comparison methods and confidence intervals. As in Figure 4, each sub-figure, sub-panels show the number of acquired cells per iteration, tSNE visualizations of using the complete data set,  visualizations using only the ActiveSVM extracted data set, and marker genes identified by ActiveSVM.}
    \label{fig:MM}
\end{figure}

To analyze ActiveSVM as a tool for the discovery of disease-specific markers, we used single-cell data from peripheral blood immune cells collected from two healthy donors and four patients who have been diagnosed with multiple myeloma (MM)\cite{chen2020dissecting}. MM is an incurable cancer of plasma cells, known as myeloma cells, that over-proliferate in the bone marrow. Although myeloma cells are typically the target of analysis because they are the causative agent of disease, peripherally circulating immune cells also contain signatures of disease, including a depleted B-cell population \cite{rawstron1998b, de2013analysis}, an increased myeloid-derived suppressor cell count \cite{malek2016myeloid}, and T-cell immunosenescence \cite{suen2016multiple, de2013analysis}.

We sought to further define transcriptional markers that distinguish healthy peripheral immune cells from the cells of MM patients. We performed feature selection using heterogeneous populations of cells labeled only by disease state. The data set contains $35159$ with $32527$ genes (Table 1). 

We compared the classification accuracy for ActiveSVM vs the other methods (Figure 5(a)(f)), and found that ActiveSVM achieved high accuracy in a limited number of steps and consistently outperformed the other methods. We tested ActiveSVM with two different cell sampling strategies, randomly sampling, and 'balanced' sampling, in which equal numbers of cells from each cell type are sampled to correct for artifacts due to different cell-type proportions between samples. We noted that although the balanced approach gave higher classification accuracy at early iterations, these differences are no longer apparent after selecting 20 genes (Figure 5(a)). 

Non-overlapping cell-type clusters were identified for healthy and MM cells in the original dataset in t-SNE projections (Figure 5(b)(h)). The non-overlapping clusters are replicated in t-SNEs constructed from $40$ genes selected using both the min-complexity strategy (Figure 5(c)(d)) and the min-cell strategy (Figure 5(i)).

Analysis of the function of the genes identified by ActiveSVM revealed most regulate house-keeping functions, suggesting that global shifts in translation and motility are disrupted in multiple myeloma patients. Translation-associated markers include Eukaryotic Translation Initiation Factor 1 (EIF1), Eukaryotic Translation Elongation Factor 1 Alpha 1 (EEF1A1), and prefoldin subunit 5 (PFDN5). Motility associated genes include ACTB, putative anti-adhesion molecule CD52, and actin-sequestering protein TMSB4X.

We also found both known and novel markers of MM within the peripheral blood immune cells. Our analysis identified TPT1, previously associated with MM \cite{ge2011quantitative}, and RACK1 (also known as GNB2L1), a scaffolding protein that coordinates critical functions including cell motility, survival and death, which is broadly upregulated in peripheral immune cells from MM patients. Although this gene has been previously associated with myeloma cells \cite{xiao2018rack1}, its regulation had not been reported in peripherally circulating immune cells. Our ability to discover MM-specific genes within peripheral immune cells suggests a broader use for discovering disease-specific genes across many different types of pathologies. 

Interestingly, the procedure also  identifies multiple members of the S100 Calcium Binding Protein Family (S100A8,S100A9 and S100A6, and S10084) \cite{xia2018s100, liu2021s100,dobreva2020single} as members of the genes sets that separate MM vs healthy samples. The S100 protein family defines a module of genes that are associated with the induction of stress response pathways. The expression of S100 genes is prognostic for a number of diseases. Specifically, a recent study found that S100A4 expression correlates with poor patient survival in mulitple myeloma and that S100A8, and S100A9 are markers that correlate with poor response of multiple myeloma patients to treatment with proteasome inhibitors and the and histone deacetylase inhibitor panobinostat \cite{liu2021s100}. The result demonstrates that ActiveSVM can automatically define groups of genes that have clinical association with disease progression and treatment outcome. The minimal gene sets generated by ActiveSVM could provide useful targeted sequencing panels for a variety of clinical tasks.

\subsection*{ActiveSVM identifies genes impacted by Cas9 based genetic perturbation}

\begin{figure}
    \centering
    \includegraphics[width=\textwidth]{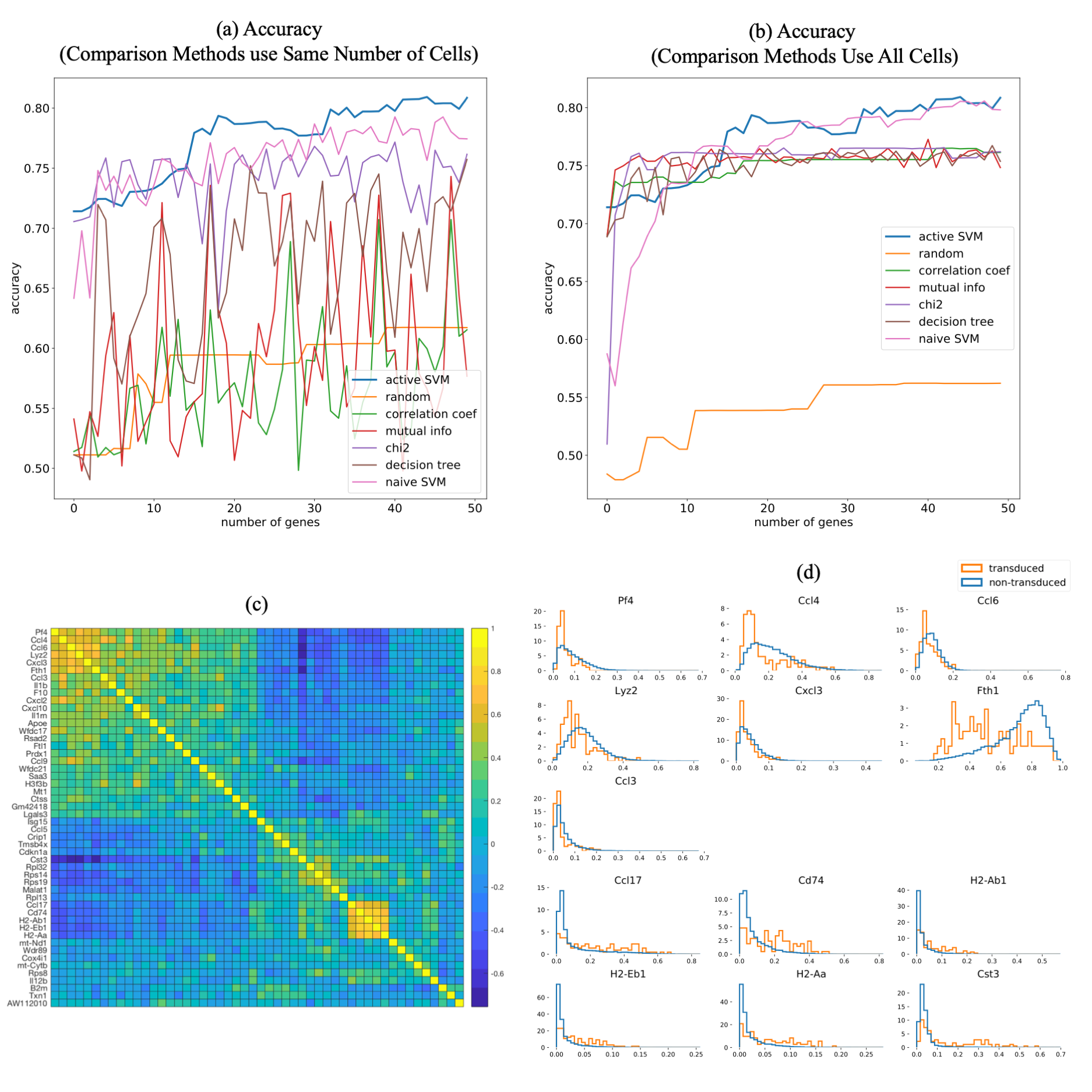}
    \caption{\textbf{Application of ActiveSVM to identify genes expression changes following Cebp knock-down with perturb-seq } The results of classification on perturb-seq data \cite{dixit2016perturb} where cells are labeled and classified as Cebp sgRNA transduced or not-transduced with a guide RNA. (a-b) accuracy of entire dataset with min-complexity strategy, where comparison methods use the same number of cells as ActiveSVM in (a) and use the entire dataset in (b). (c) correlation matrix showing pair-wise correlation coefficients for genes in Cebp perturbed cells. Correlation matrix identifies two gene modules. (d) Distributions of gene expression in Cebp sgRNA transduced (orange) or not transduced (blue) cells. Selected genes from modules in (c) shown and organized so that genes whose expression increases with Cebp perturbation are on  top and repressed genes are on the bottom of the figure.}
    \label{fig:perturb}
\end{figure}

The previous analyses above have demonstrated that ActiveSVM identifies minimal gene sets for cell-state identification across a range of single-cell mRNA-seq data sets. We, next, demonstrate that ActiveSVM provides a more general analysis tool with potential applications to a range of single-cell genomics analysis tasks. To demonstrate generalization of ActiveSVM based gene set selection across single-cell genomics tasks, we applied the method to identify marker genes in two additional applications: perturb-seq and spatial transcriptomics. 

Perturb-seq is an experimental method for performing Cas9-based genetic screens with single-cell mRNA-seq read-outs. In perturb-seq, cells are induced with libraries of guide RNA’s that target the Cas9 protein to cut and silence specific genes \cite{dixit2016perturb,replogle2020combinatorial}. Perturb-seq is performed in a pooled fashion so that a pooled set of sgRNA molecules is delivered to a cell population. Individual cells stochastically take-up specific guide RNAs, and the guide RNAs target Cas9 cuts and silences genes in the genome. Following the perturbaation experiment, single-cell mRNA-seq is applied to read both the transcriptome of each cell and the identify of the delivered sgRNA through sequencing. The advantage of the perturb-seq method is that many knock-out experiments can be performed simultaneously. However, a challenge is that noise impacts the measurement of guide RNA identify,  and, further,  the cutting of the genome by the Cas9 molecule is not complete. Due to measurement and experimental noise, identifying the impact of genetic perturbation on a cell population 
can be challenging, and various methods have been developed to boost signal \cite{replogle2020combinatorial}. We applied  ActiveSVM to identify a minimal gene set as well as down-stream effects of gene knock-down in perturb-seq data. 
 
We specifically applied ActiveSVM to analyze public data collected from mouse dendritic cells with transcription factor knock-downs \cite{dixit2016perturb}. The experiment analyzed cells in which transcription factors has been knocked-down using perturb-seq in mouse dendritic cells stimulated for 3 hours with LPS, a signal that mimics bacterial infection. 

To apply ActiveSVM to the data, we focused our analysis on knock-down of Cebp an pioneer transcription factor. We pre-processed the data to identify cells induced with sgRNA against Cebp and non-induced cells, and used transduced and non-transduced as our cell-labels. 
We applied ActiveSVM to select a minimal gene set that could classify transduced versus non-transduced cells. ActiveSVM identified minimal gene sets ($50$ genes) that achieved $~80\%$ classification accuracy on the Cebp sgRNA cell label. As we applied the class-balanced model to obtain the classification accuracy and there are only about 20 transduced cells in test set, we show the accuracy on entire dataset instead of test set. On this noisy dataset, ActiveSVM worked better than comparison methods with the condition that ActiveSVM only used a small subset of data while comparison methods performed on the entire dataset (Figure 6(b)).

We found that the discovered gene set could be decomposed into two modules of correlated genes (Figure 6c). Figure 6(c) shows a clustered correlation matrix for the $50$ identified genes. Gene expression distributions for cells in transduced vs non-transduced cells demonstrated that the modules represented two groups of genes. One group (including Pf4, Ccl4, Ccl6, Lyz2) was repressed by Cebp knock-down, and the second gene group was activated by Cebp knock-down including (Ccl17, Cd74, H2-Ab1) (Figure 6d).  

In both cases, the identified gene sets contained known targets of Cebp, the perturbed transcription factor. For example, Fth1 (ferritin, heavy polypeptide 1), Cst3, Tmsb4x, Lgals3, Ccl4, and Cd74 are all previously identified as direct binding targets of Cebp as determined by Chip-seq \cite{rouillard2016harmonizome}. Since Cebp knock-down leads to both up-regulation and down-regulation of genes, the results suggest that the factor can play both activating and repressive roles consistent with prior literature \cite{pei1990transcriptional}. 

Our analysis of the perturb-seq data set, therefore, demonstrates that ActiveSVM can be applied as a useful tool for the identification of genes modulated by perturb-seq experiments. ActiveSVM can return minimal genes sets that contain functional information. Moreover, perturb-seq has been a main application of gene targeting approaches \cite{replogle2020combinatorial}. Therefore, ActiveSVM could provide a method for identifying minimal gene sets that can be applied to increase the scale of perturb-seq data collection. 

\subsection*{ActiveSVM defines region specific markers in spatial transcriptomics data}

\begin{figure}
    \centering
    \includegraphics[width=\textwidth]{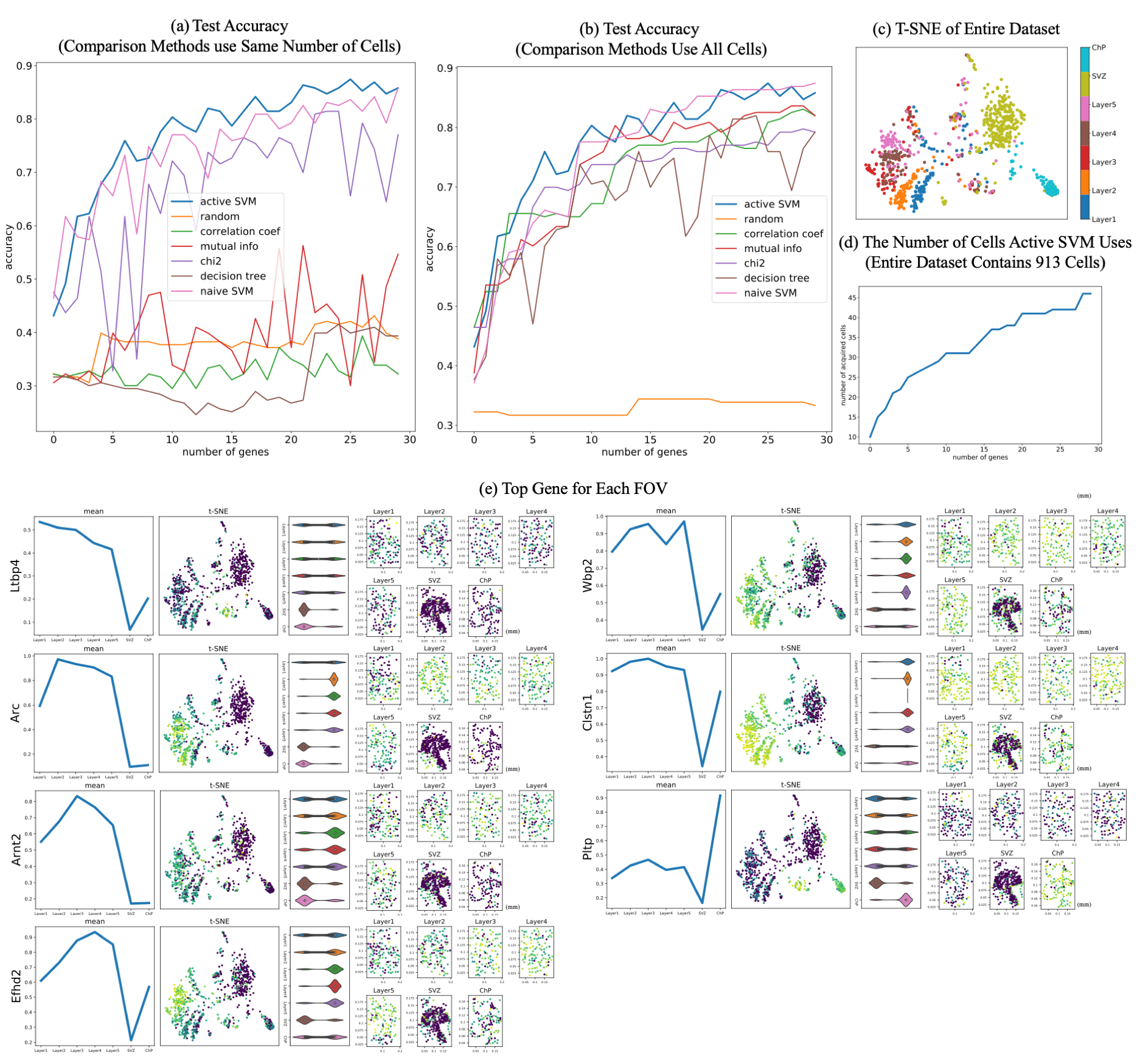}
    \caption{\textbf{Application of ActiveSVM to identify 
    region specific marker genes in the mouse brain with spatial transcriptomic data} The results of classification where cells are labeled according to fields of view (FOV) in \cite{eng2019transcriptome}. (a-b) test accuracy with min-complexity strategy, where comparison methods use the same number of cells as ActiveSVM in (a) and use the entire dataset in (b). Fields of view 1-5 correspond to 5 regions of the mouse cortex, additional fields of view are labeled SVZ (sub-ventricular zone) and ChP (chordid plexus). 
    (c) tSNE of cell transcriptomes for all cells (d) number of cells used per iteration (e) Sample of identified genes where each sub-panel shows mean  expression across FOV/brain regions for selected gene, a tSNE plot colored by expression of selected gene,   a violin plot of single cell gene expression values for selected gene in FOV/brain region, and spatial plots of each field of view where dots represents cells in 2D imaging slice, cells are colored by intensity of selected gene and units are in millimeters.}
    \label{fig:seqFish}
\end{figure}

Finally, to further demonstrate the generality of the ActiveSVM approach, we applied the procedure to identify minimal gene sets for classification of cells by spatial location in spatial transcriptomics data. Spatial transcriptomics is an emerging method for measuring mRNA expression within single cells while retaining spatial information and cellular proximity within a tissue. As an example, in SeqFish+, an imaging based spatial transcriptomics method, cells are imaged in their tissue environment, and mRNA transcripts are counted using single-molecule imaging of mRNA spots \cite{eng2019transcriptome}. In all spatial transcriptomics applications, a common goal is the identification of genes that mark specific spatial locations within a tissue sample. Additionally, spatial imaging methods are commonly limited by imaging time. While Seqfish+ can profile $10,000$ mRNA molecules per cell, the identification of reduced gene sets would reduce imaging time and throughput. 

We applied ActiveSVM to identify genes associated with specific spatial locations in the mouse brain. We used a seqFISH+ data set in which the authors profile $10,000$ mRNA molecules in 7 fields of view (FOV) in the mouse brain \cite{eng2019transcriptome}. Fields of view correspond to spatially distinct regions of the mouse cortex as well as the sub-ventricular zone and chordid plexus. We used the spatial location labels provided by [\cite{eng2019transcriptome}] to identify seven different brain locations (Fields of view 1-5 corresponding to Cortex Layers 2/3 through Layer 6; FOV 6 is sub-ventricular zone, and FOV 7 is chordid plexus). Applying the spatial location labels as class labels, we applied ActiveSVM to identify genes that could allow classification of single-cells by their location in one of the seven classes and to define marker genes that correspond to specific spatial locations. 

We identified gene sets of $<30$ genes that enabled location classification with greater than $85\%$ accuracy with min-complexity strategy (Figure 7(a)). ActiveSVM used only $10$ cell at each iteration but worked better than comparison methods who performed on the entire dataset (Figure 7(b)).

In the spatial application, the result means that the $\sim 30$ genes are sufficient to classify single-cells as belonging to one of the $7$ spatial classes. In Figure 7, we show the mean expression of identified genes across cortical fields of view corresponding to a sweep through cortical layers 2/3 through 6 as well as SVZ and CP. Our analysis identifies markers Prex1 that are specific to the upper cortical layers of the brain.  Efhd2, a calcium binding protein linked to Alzheimer's disease and dementia, was similarly expressed in lower cortical layers \cite{vega2016efhd2,borger2014calcium}. Finally, Pltp, a Phospholipid transfer protein, was localized to the chordid plexus. In Figure 7e, we show the spatial distribution of these genes including their mean expression across regions, violin plots documenting expression distribution, and renderings of the single-cells within the field of view and the relative expression of each gene.   

The spatial analysis demonstrates that a broad range of different experimental variables can be applied as labels. In each case ActiveSVM discovers genes that allow classification of cells according to labels and identifies interesting genes. Regional gene marker identification is a major task in seqFish data analysis and ActiveSVM is able to identify genes enriched in different brain regions automatically. Such spatial information could provide interesting new insights into disease processes mediated by genes like Efhd2. 

\section*{Discussion}
In this paper, we introduce ActiveSVM as a feature selection procedure for discovering minimal gene sets in large single-cell mRNA-seq datasets. ActiveSVM extracts minimal gene sets through an iterative cell-state classification strategy. At each round the algorithm applies the current gene set to identify cells that classify poorly. Through analysis of misclassified cells, the algorithm identifies maximally informative genes to incorporate into the target gene set. The iterative,active strategy reduces memory and computational costs by focusing resources on a highly informative subset of cells within a larger data set. By focusing computational resources on misclassified cells, the method can run on large data sets with more than one million cells. We demonstrate that ActiveSVM is able to identify compact gene sets with tens to hundreds of genes that still enable highly accurate cell-type classification.   We demonstrate that the method can be applied to a variety of different types of data set and single-cell analysis tasks including perturb-seq data analysis and spatial transcriptomic marker gene analysis. 

Conceptually, we refer to our strategy `active' because it actively selects data examples (here cells) at each iteration for detailed analysis . Our algorithm specifically selects cells that within the margin of the SVM classifier, and uses these poorly classified cells to search for maximally informative genes (features). In traditional active learning strategies, an algorithm is typically called active when it can directly query an oracle for data examples that meet a criteria \cite{settles2009active,settles2011theories}. In the tradition of active learning, our ActiveSVM procedure queries the SVM classifier for cells that have been misclassified, and then expends computational resources to analyze all genes within that limited subset of cells to discover informative genes. Thus, while our algorithm cannot query the biological system directly for cells that meet a specific criteria, the algorithm queries the data set itself for informative examples, and therefore we refer to it as `active'.  Our current work focuses on a single classification method, the support vector machine, as the computational engine. Active learning methods  can be applied more broadly to additional classification strategies like neural network based classification as well as to additional types of analysis like data clustering and gene regulatory network inference. 

Our method also has some conceptual similarity to boosting methods \cite{schapire2003boosting, schapire1999brief}. Boosting algorithms (e.g AdaBoost) train a series of 'weak' learners for a classification tasks, and then combine these weak classifiers to generate a strong classifier. In boosting a single weak learner may initially obtain moderate performance on a task. The performance of weak learners is improved through iterative training of additional learners and focusing their training on difficult data examples, for example,  misclassified examples. The boosting algorithm constructs a final, strong classifier by combining the results of the ensemble of weak classifiers through a weighted majority vote. Our method is distinct from conventional boosting, because we search for a minimal set of features in our data that allows a single SVM classifier to achieve high-accuracy classification. However, ActiveSVM feature selection shares conceptual ideas with boosting in that both methods focus analysis on challenging examples and combine information to achieve strong classification from initially weak classifiers. 

ActiveSVM provides an iterative strategy for extracting a compact set of highly informative genes from large single cell data sets. Biologically, recent work highlights the presence of low-dimensional structure within the transcriptome \cite{heimberg2016low}. Low-dimensional structure emerges in gene expression data because cells modulate their physiological state through gene expression programs or modules that contain large groups of genes. Since genes within transcriptional modules have highly correlated expression, measurements performed on a small number of highly informative signature genes can be sufficient to infer the state of a cell \cite{cleary2017efficient}.  Low-dimensional structure can be exploited to decrease measurement and analysis costs since a small fraction of the transcriptome must be measured to infer cellular state. We developed ActiveSVM as a scalable strategy for extracting high information content genes within a sharply defined task, cell-state classification. 

In ActiveSVM we apply an active learning strategy to reduce the computational and memory requirements for analyzing single-cell data sets by focusing computational resources on 'difficult to classify' cells. In the future, active learning strategies could be applied directly at the point of measurement. In genomics measurement resources often limit the scale of data acquisition.  In future work we aim to develop strategies that can improve the on-line acquisition of single-cell data. Active strategies could be implemented at the point of measurement by only sequencing or imaging the content of cells that meet a criteria. Even more broadly, it might be possible to induce a biological system to generate highly informative examples through designed experimental perturbation \cite{jiang2019active}.

\section*{Data Availability}
All data used in the paper has been previously published. 
The PBMC Single-cell RNA-seq data have been deposited in the Short Read Archive under accession number SRP073767 by the authors of \cite{zheng2017massively}. Data are also available at \url{ http://support.10xgenomics.com/single-cell/datasets}.

The original Tabula Muris dataset is available at \url{https://figshare.com/projects/Tabula_Muris_Transcriptomic_characterization_of_20_organs_and_tissues_from_Mus_musculus_at_single_cell_resolution/27733}. 

The original multiple myeloma PBMC data, containing 2 healthy donors and 4 multiple myeloma donors, is available at \url{https://figshare.com/articles/dataset/PopAlign_Data/11837097/3}.

The 10x genomics Megacell data set is available at \url{ http://support.10xgenomics.com/single-cell/datasets}.

The perturb-seq data set \cite{dixit2016perturb} is availble at \url{https://www.ncbi.nlm.nih.gov/geo/query/acc.cgi?acc=GSM2396856}

The spatial transcriptomics data \cite{eng2019transcriptome} is available \url{https://github.com/CaiGroup/seqFISH-PLUS.} 

\section*{Code Availability}

Our method is integrated as a install-able Python package called activeSVC. The installation instructions and user guidance are shown at \url{https://pypi.org/project/activeSVC}.  The source codes of activeSVC and some demo examples are publicly available on GitHub at \url{https://github.com/xqchen/activeSVC}.

The Python package provides six callable functions: (1) $min\_complexity$; (2) $min\_acquisition$, the min-cell strategy; (3) $min\_complexity\_cv$, which use cross validation\cite{arlot2010survey} and grid-search\cite{syarif2016svm} to train the best SVM estimator at each iteration; (4) $min\_acquisition\_cv$, the min-cell strategy with cross validation and grid-search; (5) $min\_complexity\_h5py$, for large h5py\cite{h5py} data files, it only loads the part of data, the rows and columns of selected genes and cells, instead of loading the entire dataset into memory; (4) $min\_acquisition\_h5py$, is similar with $min\_complexity\_h5py$ but uses min-cell strategy. All include the algorithm for both randomly and 'balanced' sampling. We implement the SVM classifier with the LinearSVC package from scikit-learn\cite{scikit-learn} library, which is implemented in term of LIBLINEAR\cite{fan2008liblinear}. And we use parfor\cite{parfor} package to parallelize for-loops to accelerate algorithm for large datasets. There are three hyper-parameters to set: balance (boolean), num\_features (int),and num\_samples (int), to identify the sampling strategy, the number of genes to select, and the number of cells each iteration. 

In the GitHub project, we use the PBMC dataset\cite{zheng2017massively} and Tabula Muris dataset\cite{tabula2018single} as examples to show the procedure and its  performance of $min\_complexity$ and $min\_acquisition$. We also have the test examples of $min\_complexity\_cv$ and $min\_acquisition\_cv$ on PBMC dataset and the demo projects of $min\_complexity\_h5py$ and $min\_acquisition\_h5py$ on 1.3 millions mouse brain `megacell' dataset\cite{genomics20171}. The notebooks contain downloading dataset, preprocessing, and selecting genes with our method. Besides, we created Google Colaboratory project for these two examples that PBMC demo is at \url{https://colab.research.google.com/drive/16h8hsnJ3ukTWAPnCB581dwj-nN5oopyM?usp=sharing}, Tabula Muris demo is at \url{https://colab.research.google.com/drive/1SLehIKIQqpjK6BzEKc9m0y3uJ_LBqRzA?usp=sharing}, and PBMC cross-validation demo is at \url{https://colab.research.google.com/drive/1fhQ8GD3NyzB3w0vof9WimXK6BLqDNuDC?usp=sharing}.

\section*{Experiments}
\subsection*{Data Pre-processing}
We found that ActiveSVM was able to achieve high-performance across pre-processing strategies. For single-cell mRNA-seq column normalization (l2) was important for removing artifacts due to cell to cell variability in mRNA capture. However, additional algorithm was not sensitive to additional pre-processing steps. 

\subsubsection*{PBMC, Tabula Muris, Multiple-Myeloma}
These three data sets were pre-processed for a prior publication \cite{chen2020dissecting} via column normalizaiton.  In each experiment, we removed the columns and rows where all values are zero. Then, gene expression matrices were first columns normalized and log transformed. For a cell $j$, each gene $x_{ij}$ (gene $i$ in cell $j$) is first normalized as $\tilde{g}_{i j} = \frac{g_{i j}}{\sum^n_{i=1} g_{i j}} $ where $n$ is the number of genes in the transcriptome. 


\subsubsection*{Mega-cell data set, perturb-seq, spatial transcritomics}
For these data sets, we removed the columns and rows where all values are zero. Then we did $l^2$-normalization along each cell to scale input cell vectors individually to unit squared-norm.

\subsection*{Parameters}
Here we provide the algorithm parameters we used for ActiveSVM in Table 2,3,4. Besides the training set and test set, there are 15 user-defined hyper-parameters in ActiveSVM, five of which are about the feature selection procedure and the other ten are commonly-used parameters for linear SVM classifier. The detailed description about all parameters of ActiveSVM are detailed described in the integrated package page \url{https://pypi.org/project/activeSVC/}.

As for comparison methods, correlation coefficient, mutual information, and chi-squared methods don't have specific parameters to set. We implemented them with scikit-learn\cite{scikit-learn} package 'SelectKBest'. For feature importance scores from decision tree and naive SVM, we did grid-search on key parameters based on 3-fold cross validation at each step. The parameters of decision tree are $criterion$ and $min\_samples\_leaf$ and of naive SVM are $tol$ and $C$. 

\begin{table}[!htbp]
\centering
\caption{Parameters of ActiveSVM (PBMC and mouse megacell datasets).}
\begin{tabular}{*6c}
\toprule
{} & PBMC & PBMC& mouse megacell & mouse megacell \\
{} & (min-complexity) & (min-cell) & (min-complexity) & (min-cell) \\
\midrule
$num\_features$ & 50 & 20 & 50 & 50\\
$num\_samples$ & 20 & 100 & 20 & 100\\
$init\_features$ & 1 & 1 & 1 & 1\\
$init\_samples$ & 20 & 200 & 20 & 100\\
$balance$ & True/False & True & True & True\\
$penalty$ & 'l2' & 'l2' & 'l2' & 'l2'\\
$loss$ & squared\_hinge & squared\_hinge & squared\_hinge & squared\_hinge\\
$dual$ & True & True & True & True\\
$tol$ & 1e-4 & 1e-4 & 1e-4 & 1e-4\\
$C$ & 1.0 & 1.0 & 1.0 & 1.0\\
$fit\_intercept$ & True & True & True & True\\
$intercept\_scaling$ & 1 & 1 & 1 & 1\\
$class\_weight$ & None & None & 'balanced' & 'balanced'\\
$random\_state$ & None & None & None & None\\
$max\_iter$ & 1000 & 1000 & 1000 & 1000\\
\bottomrule
\end{tabular}
\end{table}

\begin{table}[!htbp]
\centering
\caption{Parameters of ActiveSVM (Tabula Muris and MM datasets).}
\begin{tabular}{*5c}
\toprule
{} & Tabula Muris & Tabula Muris & MM & MM \\
{} & (min-complexity) & (min-cell) & (min-complexity) & (min-cell) \\
\midrule
$num\_features$ & 150 & 500 & 40 & 40\\
$num\_samples$ & 20 & 200 & 20 & 100\\
$init\_features$ & 1 & 1 & 1 & 1\\
$init\_samples$ & 20 & 200 & 20 & 100\\
$balance$ & True/False & False & True/False & False\\
$penalty$ & 'l2' & 'l2' & 'l2' & 'l2'\\
$loss$ & squared\_hinge & squared\_hinge & squared\_hinge & squared\_hinge\\
$dual$ & True & True & True & True\\
$tol$ & 1e-4 & 1e-4 & 1e-4 & 1e-4\\
$C$ & 1.0 & 1.0 & 1.0 & 1.0\\
$fit\_intercept$ & True & True & True & True\\
$intercept\_scaling$ & 1 & 1 & 1 & 1\\
$class\_weight$ & None & None & None & None\\
$random\_state$ & None & None & None & None\\
$max\_iter$ & 1000 & 1000 & 1000 & 1000\\
\bottomrule
\end{tabular}
\end{table}

\begin{table}[!htbp]
\centering
\caption{Parameters of ActiveSVM (perturb-seq and seqFish datasets).}
\begin{tabular}{*3c}
\toprule
{} & perturb-seq & seqFish \\
{} & (min-complexity) & (min-complexity) \\
\midrule
$num\_features$ & 50 & 30\\
$num\_samples$ & 500 & 10\\
$init\_features$ & 1 & 1 \\
$init\_samples$ & 1000 & 10 \\
$balance$ & True & False \\
$penalty$ & 'l2' & 'l2' \\
$loss$ & squared\_hinge & squared\_hinge\\
$dual$ & True & True \\
$tol$ & 1e-6 & 1 \\
$C$ & 1.0 & 10\\
$fit\_intercept$ & True & True\\
$intercept\_scaling$ & 1 & 1\\
$class\_weight$ & 'balanced' & None\\
$random\_state$ & None & None \\
$max\_iter$ & 1,000,000 & 100,000\\
\bottomrule
\end{tabular}
\end{table}

\subsection*{Confidence Intervals}
Confidence intervals were estimated using a proportion confidence interval\cite{brown2001interval} as  $\text{interval} = z \sqrt{\frac{\epsilon * (1 - \epsilon)}{n})}$ where $z = 1.96$ for $95\%$ confidence and $n$ is the number of cells and $\epsilon$ the observed error.

\bibliography{bibliography}

\bibliographystyle{ieeetr}

\end{document}